\documentclass[twocolumn,showpacs,preprintnumbers,amsmath,amssymb]{revtex4}

\usepackage{graphicx}
\usepackage{dcolumn}
\usepackage{bm}

\begin{document}
\def\al{&\!\!\!\!}
\def\x{{\bf x}}
\def\f{\frac}
\preprint{APS/123-QED}

\title{Dark companion of baryonic matter}
\author{Yousef Sobouti}
\email{sobouti@iasbs.ac.ir}
\affiliation{Institute For Advanced Studies in Basic Sciences-
Zanjan\\P.O. Box 45195-1159, Zanjan 45195, Iran
}%


\date{\today}

\begin{abstract}
Whenever and wherever one talks of dark matter, one does so when
and where there is a luminous matter and a dynamical issue to be
settled. We promote this observation to the status of an axiom and
assume that there is a dark companion to every luminous matter and
there are orders to this companionship. To pursue the proposition
in a formal and quantitative manner, we consider the anomalous
rotation curves of spiral galaxies. From the available
observations, we infer the gravitational potential prevailing in
the outer parts of the galaxy and, thereof, construct the $tt$-
component of the metric of the embedding spacetime. Next we
examine a perfect fluid candidate as the dark companion and solve
the relevant GR equations. We are able to determine the strength
and the distribution of the dark fluid that accompanies a point
baryonic mass. Finally, we argue that the whole paradigm can be
explained just as well in terms of an alternative theory.\\

\end{abstract}

 \pacs{90}                            

\maketitle

\section{\label{sec:level1}Introduction}

That the baryonic matter of galaxies, clusters of
galaxies, or for that matter, of the universe at large, does not
provide sufficient gravitation to explain the observed dynamics of
the systems, is an established fact. To solve the dilemma,
dark matter/energy scenarios and/or alternative theories of
gravitation have been speculated and debated. The fact, however,
remains that the proponents of dark matter/energy have always looked for it in
baryonic environments. No one has, so far, reported a case where
there is no luminous matter and/or cosmic radiation, but there is a
dynamical problem to be solved. In  view of this negative
observation, it is not unreasonable to hypothesize that any
luminous matter has a dark companion and there are rules to this companionship  as regards
the magnitudes and the distributions of the  companion and the matter itself.

On the other hand such a point of view, that denies the independent
existence of the dark matter/energy, is equivalent to the
assumption that the known theories of gravitation , based on
baryonic matter alone, do not tell the whole  story and there is room for
amendments. This conclusion, in turn, reduces the distinction
between the dark matter/energy hypotheses and alternative theories  to the  level
 of semantics.  As long as the dark matter displays no other physical
 characteristics than gravitation, one will have the
option, either to assume a dark  component to every baryonic
matter subject to certain rules, and account for its gravitation
in a conventional  way, or simply adhere to the baryonic matter
but resort to an alternative law of gravitation. The two points of
view should be equivalent.

 With this perspective in sight, here we
confine the discussion to the problem of spiral galaxies. There is
substantial amount of information in the observed rotation curves
of spirals to construct an empirical law of gravity. This step
leads to a partial construction of the spacetime metric around the
galaxy. Next we look for the required modification of the field
equations of GR to ensure the self-consistency of the theory. The
end results can then be interpreted, interchangeably, either in
terms of a dark matter scenario or in terms of an alternative
theory of gravitation

\section{\label{sec:level1}Observed facts and implications}
 There are three main characteristics to the rotation curves of spirals.
\begin{itemize}

\item a) They often
have a flat asymptote at far distances from the galaxy, see e.g.
\cite{sho}, \cite{rw},  \cite{sg}, \cite{sv}, $\&$ \cite{bbs}.

\item b) Their asymptotic speed is, more often than not, proportional to the
fourth root of the mass of the galaxy, the Tully-Fisher relation \cite{tf}.

\item c) Deviations from the classical concepts (in this case
gravitation) show up in large scale systems and at large distances.
\end{itemize}
These observed facts will be treated as axioms and will serve as the
starting point of what follows.\\

\textbf{The model}: A test
object orbits a galaxy at far distances from it, and has a constant
distance-independent circular speed, item \emph{a} above. To have such a speed one
requires a force field that fades away as $r^{-1}$ and, therefore, a
gravitational potential as $\ln r$. In the GR perspective, the metric
field surrounding the galaxy should also exhibit the same
logarithmic behavior. \\

Let us view the galaxy from afar and
approximate it by a point mass. The spacetime around will accordingly be
spherically symmetric and isotropic:
 \begin{eqnarray}
 ds^2 = -B(r) dt^2 + A(r) dr^2 + r^2 d\Omega^2.
 \end{eqnarray}
It is customary to write $B(r)=1+2\phi(r)/c^2$ and, in the weak field
regime, to consider $\phi(r)$ as the gravitational potential.
One, however, knows that at close distances the gravitation is newtonian,
item \emph{c} above, and at far distances as we just learned
should be proportional to $\ln r$. Thus, we let
 \begin{eqnarray}
 B(r)=1-\frac{s}{r}+\lambda \ln r,
 \end{eqnarray}
where $s=2G M/c^{2}$ is the Schwarzschild radius of the galaxy,
and $\lambda$ is a  dimensionless constant.  It will emerge as part of the sought-after
modification to the field equations.\\

\textbf{What is $\lambda $?} The answer is in the Tully-Fisher
relation, item \emph{b} above. From Eq. (2), the circular speed of an orbiting test
object is
\begin{eqnarray}
v^{2}=\frac{1}{2}c^{2}r\frac{d
B}{dr}=\frac{GM}{r}+\frac{1}{2}\lambda c^{2}.
 \end{eqnarray}
By Tully-Fisher, the asymptotic speed, $\sqrt{\lambda c^{2}/2}$,
is proportional to the fourth root of the mass of the galaxy. This
implies
\begin{eqnarray}
\lambda=\lambda_{0}\left(\frac{M}{M_{\odot}}\right)^{1/2}.
\end{eqnarray}\\

\textbf{What is} $\lambda_0$ \textbf{?} In his theory of MOND,
Milgrom \cite{mil} proposes a law of gravitation whose strong  and
weak limits are  the newtonian gravity, $g_N = GM/r^{2}$, and
$(a_0 g_N)^{1/2}$, respectively. From the inspection of the
observed data Begeman et al. \cite{bbs} find
 $a_0\approx1.2\times10^{-8}$ $\rm{cm ~sec^{-2}}$.
With $\lambda$ of Eq. (4), Eq. (3) has the same strong and weak limits
  of MOND. We use this coincidence  to find $\lambda_0$. We divide Eq. (3) by $r$,
  substitute for $\lambda$ from Eq. (4), and identify the resulting term,
  $\frac{1}{2}\lambda_{0}c^2({GM/r^2GM_{\odot}})^{1/2}$ with Milgrom's $a_0g_N$. We obtain
\begin{eqnarray}
\lambda_0 = \left[\f{4a_0}{c^2} \f{GM_\odot}{c^2}\right]^{1/2} \approx 2.8\times 10^{-12}.
\end{eqnarray}
The problem is partially solved. We have constructed an empirical
law of gravity, whose strong and weak limits are those of MOND.
There remains to build  the empirically constructed $B(r)$ into a consistent general relativistic formalism. \\

\section{\label{sec:level1}The modified field equations}
We seek this modification by adding a new tensor term to Einstein's field equations.
To respect the Biancci identities and the conservation laws of the baryonic matter,
this tensor should have a vanishing covariant divergence.This is best achieved by
adopting a dark matter point of view. We assume the galaxy, approximated by a point mass, $M$,
has a `dark perfect fluid' companion, with the energy momentum tensor
\begin{eqnarray}
\al\al T_d^{\mu\nu} = p_d g^{\mu\nu} + (\rho_d +p_d)U_d^\mu U_d^\nu,\\\cr
\al\al {T_d^{\mu\nu}}_{;\nu} = 0,~~~U_d^t = g^{tt},~~U_d^i = 0,
\end{eqnarray}
where $\rho_d$ and $p_d$ are the density and the pressure  of the dark fluid, respectively.
The fluid is  spherically symmetric and is at rest. Its 4-velocity, $U_d^\kappa$, is indicated in Eq. (7).
The amended field Equations in the `baryonic vacuum' of the galaxy now reads
\begin{eqnarray}
R_{\mu\nu}- \frac{1}{2}g_{\mu\nu}R=-8\pi G T_{d\mu\nu}.
\end{eqnarray}
The baryonic matter of the galaxy has a $\delta$-function density distribution
and is of zero pressure. It will show up as a constant of integration in the final stage of integrations.
The spacetime metric is still spherically  symmetric and isotropic as in Eq.(1).
From Eq. (8), the two combinations,  $R_{tt}/2B + R_{rr}/2A + R_{\theta\theta}/r^{2}$ and
$R_{tt}/B+R_{rr}/A$, give
\begin{eqnarray}
\al\al\frac{d}{dr}(\frac{r}{A})=1-8\pi G \rho_{d}r^{2},\\\cr
\al\al\frac{B'}{B}+ \frac{A'}{A}=8 \pi G r A(\rho_{d}+ p_{d}).
\end{eqnarray}
To this we add the time-component of Eq.(7),
\begin{eqnarray}
\frac{p'_{d}}{\rho_{d}+\rho_{d}}=-\frac{1}{2}\frac{B'}{B}.
\end{eqnarray}
The space components of Eq. (7) are trivially satisfied on  account of $U_d^i = 0.$
Equation (9) immediately integrates into
\begin{eqnarray}
\frac{1}{A}=1- \frac{s}{r}- 2 G \frac{m_{d}(r)}{r}, ~~m_{d}(r)=4\pi\int\rho_{d} r^{2}dr,
\end{eqnarray}
where $s$ is the integration constant and as in Eq. (2) should be
identified with Schwarzschild's radius of the galaxy. We already
have inferred $B(r)$ from observations, Eq. (2). There remains to
solve Eqs. (10)-(12) for $ A(r), \rho_{d}$ and $p_{d}$. Exact
solutions are probably to be obtained numerically. Their weak
field approximations are, however, analytically available and are
inspiring. We consider the dimensionless quantities $\lambda$,
$s/r$, $\&$ $G m_{d}(r)/r$  much smaller than 1, and keep only
their first order terms in all calculations. We also assume and
verify later that $p_{d}\ll \rho_{d}$. With these provisions, Eqs.
(10) and (12) gives
\begin{eqnarray}
\al\al m_{d}(r)=\frac{\lambda}{2G}r,~~~~~
\rho_{d}(r) = \frac{\lambda}{8 \pi G}\frac{1}{r^{2}}.
\end{eqnarray}
With this $\rho_{d}$, Eq. (11) integrates into
\begin{eqnarray}
p_{d}(r) \al=\al \frac{\lambda}{32\pi G}\left[\frac{\lambda}{r^{2}}+ \frac{2}{3}\frac{s}{r^{3}}\right]\nonumber\\\cr
\al=\al\frac{\lambda}{4}\rho_{d}+\frac{(2\pi G \lambda)^{1/2}}{3}s\rho_{d}^{3/2}\ll\rho_d.
\end{eqnarray}
The second equality in Eq. (14) is obtained by eliminating $r$
between $\rho_{d}(r)$ and $p_{d}(r)$. The inequality is on account
of the smallness of $\lambda$. If one is allowed to use the
terminologies and concepts of the real world's physics,
one might say the dark fluid has a barotropic equation of states.\\

The goal set in the introduction is, at least partially, arrived at.
For a point mass $M$ (Schwarzschild's radius $s$) we have found a static dark fluid companion.
Its strength and distribution is given by Eqs. (13), (14), (4), and (5).
This parlance, however, is no more than borrowing a jargon from the physics of the
observable world to explain the purpose. Equivalently, one may choose to say that the gravitation
produced by a point mass is not newtonian and there is a logarithmic correction to it.
Or, rather, the spacetime around a point mass is not that of Schwarzschild but that given by Eqs. (2), (12) and (13).\\

\textbf{How much dark matter in a typical spiral?} The question
should be qualified by giving the radius inside which the mass is
inquired.   From Eq. (13), after inserting a factor $c^2$, which
so far was suppressed, and substituting for $\lambda$ from Eqs.
(4) and (5), one finds
   \begin{eqnarray}
   \f{m_d(r)}{10^{10}M_\odot} \al=\al 2.8\left[\f{M}{10^{10}M_\odot}\right]^{1/2}\left[\f{r}{10~kpc}\right],~~~\rm{or}\cr
   \f{m_d(r)}{M_\odot} \al=\al 1.4\times
   10^{-4}\left[\f{M}{m_\odot}\right]^{1/2}\left[\f{r}{a.u.}\right].
   \end{eqnarray}
The dark matter inside a sphere, centered on the baryonic point
mass, is proportional to the radius of the sphere and to square
root of the mass residing at the center. For the Milky Way of
total stellar + \emph{HI} mass $\approx 6\times 10^{10}M_\odot$ at
$r= 10$ and 50 kpc (the later is the distance to the Large
Magellanic Clouds) the dark matter is 7 and 35 $\times 10^{10}
M_\odot$, respectively.  They amount to 55\% and 83\% of the
required dynamical mass. The dark matter accompanying Sun within
the outermost reaches of the solar system, 100 a.u., say, is
$\approx 1.4 \times 10^{-2}M_\odot$ and less by a factor of one
hundred at Earth's distance. \\

\textbf{Spacetime is not flat.} From Eqs. (2), (12), and (13), one
has
 \begin{eqnarray}
\f{1}{A} = 1-\lambda - \f{s}{r},~~~B=1+\lambda\ln r - \f{s}{r}.
 \end{eqnarray}
 Contracting Eqs. (8) and (6) and using Eqs. (13), and (16), gives the scalar curvature of the 4-spacetime
 \begin{eqnarray}
 R =8\pi G\left[3p_d-\rho_d\right]\approx -\f{\lambda}{r^2}.
 \end{eqnarray}
 The scalar curvature of the 3-space, calculated from the 3-space metric, $g_{ij},~i,~j=r,~\theta,~\varphi$, turns out to be
 \begin{eqnarray}
 R^{(3)} \approx -2\f{\lambda}{r^2}.
 \end{eqnarray}
 Both curvatures are negative and fade away as $r^{-2}$.
 This is to be contrasted with  Schwarzschild's spacetime, where the spacetime  and the 3-space have zero scalar curvatures.\\

\textbf{There is an excess lensing.} This is to be expected on
account of the excess gravitation of the dark companion. A light
ray impinging on a lens from infinity and escaping  to infinity
bends by an  angle \cite{wein}
\begin {eqnarray}
\beta = 2\int_{r_0}^\infty A^{1/2}\left[ \left(\f{r}{r_0}\right)^2\f{B(r)}{B(r_0)}-1\right]^{-1/2}
\f{dr}{r^2} - \pi,
\end {eqnarray}
where $r_0$ is the distance of the closest approach to the lens.
Substituting for $A$ and $B$ from Eq. (17), and keeping only the first
order terms in $s/r$, $s/r_0$ and $\lambda$ in the integral, gives
\begin {eqnarray}
\beta = 2 \f{s}{r_0} + \f{1}{2}\pi \lambda =  2 \f{s}{r_0} +
\f{1}{2}\pi \lambda_0\left( \f{M}{M_\odot}\right)^{1/2}.
\end {eqnarray}
In this first order approximation, the excess deflection,
$\f{1}{2}\pi \lambda$, is independent of the impact parameter of
the incident light ray. It is proportional to the square root of
the mass of the lens, and could be large in large systems,
clusters of galaxies, say. \\

\textbf{Solar system implications:} From Eq. (20), for a light ray
grazing Sun's limb, the excess deflection amounts to $\approx
10^{-6}$ arcsec, negligible  compared with the general relativistic value of 1.8 arcsec.\\
 Precession of the perihelion of an orbit is  obtained from \cite{wein}
 \begin {eqnarray}
\delta\phi\al=\al
2\int_{r_-}^{r_+}\f{A^{1/2}(r)}{J^2}\left[\f{1}{B(r)}-E-\f{1}{r^2}\right]^{-1/2}\f{dr}{r^2}
-2\pi\nonumber\cr\\
\al=\al
3\pi\f{s}{L}+\lambda(\f{2s}{L})^{1/2}\left[1+\f{3}{4}e\right],
 \end {eqnarray}
 where $r_\pm$ are the aphelion and the perihelion of the
 orbit, and $E , J, L,$ \& $e$ are its energy, angular momentum, semi latus rectum,
 \& eccentricity, respectively. Again in view of the smallness of $\lambda$,
the excess precession is much smaller than the conventional GR value.\\

\textbf{Beyond the point mass}: The metric coefficient of Eq. (16)
are for a point mass. Galaxies at close and intermediate distances
do not appear as such. In view of the smallness of $\lambda$ and
the proportionality of $m_d(r)$ to $r$, however, contributions of
$\lambda$ terms are significant only at far distances. Otherwise
the gravitational potential is essentially  newtonian and the
spacetime as that of GR, item \emph{c} above. Thus, considering
the present-day accuracies of the observational data, one may
generalize Eqs. (16) by replacing the point mass term, $-s/r$ , by
the whatever GR requires for an extended object, and leave the
$\lambda$-terms as they are. In the weak field regime, the metric
coefficients become
\begin {eqnarray}
\f{1}{A} \al=\al 1-\lambda -2Gc^{-2}\int \textbf{dr}'^3 \rho(\textbf{r}')|\textbf{r-r}'|^{-1},\\
B\al=\al1+\lambda\ln r -2Gc^{-2}\int \textbf{dr}'^3
\rho(\textbf{r}')|\textbf{r-r}'|^{-1}.
\end {eqnarray}
It should, however, be noted that
this generalization does not follow from a founding principle. It can only
serve practical exigencies.
\\

 \textbf{Kinship with $f(R)$ gravity of  \cite{so}}:
 In  \cite{so}
 we introduce an Einstein-Hilbert action, which is essentially
 $R^{1-\lambda/2} \approx R[1- \f{1}{2}\lambda\ln r]$, and obtain a spacetime metric with a logarithmic
 correction to it. In the weak field regime, its near- and far- distance limits
 are the same as  those of the present paper (and of those of MOND).
 For practical purposes the two theories are identical. There is, however,
 an axiomatic advantage to the present formalism.
 As noted earlier, the present formalism respects the Biancci identities and the conservation laws
 of the baryonic matter. No $f(R)$ formalism does so.

 Mendoza et al \cite{men} shows that, in the spacetime of \cite{so}, the light
 and the gravitational waves propagate with the speed of light in vacuum. Their conclusion
 is also true in the present case, on account of the identical near- and far- distance
 limits of the present and the $f(R)$ formalisms of \cite{so}.

\section{ Concluding remarks}
The proposed formalism is a modified GR paradigm or, equivalently,
a dark matter scenario, to understand the anomalous rotation
curves of the spiral galaxies. It is an inverse approach. From the
available observations, the gravitational potential and, thereof,
part of the spacetime metric is constructed. Next the GR formalism
is called upon to infer what modifications to Einstein's field
equations produces a cohesive and self-consistent picture.
Naturally, the credibility of the proposition depends on how
accurately the axiomatized model of section II describes the
realities of the skies. For example, if the future observations
reveal a decline in the rotation curves at very far distances, as
some authors have pointed at such indications in the observed data
\cite{pvkh} or entertained it on theoretical grounds \cite{mof},
the model and the empirically inferred gravitational potential
should be adjusted accordingly.\\

Ascription of a dark companion to every baryonic matter, the rules
of the companionship, and the consequent equivalence of the whole
scenario to an alternative GR theory are the highlights of the
paper.  However, only the case of a point mass is handled. An
axiomatic generalization to many body systems and continuous
distributions of luminous matter requires further deliberations
and better and more extensive observational data. One might need
further postulates. The difficulty lies in the fact that a) there
is no superposition principle to resort to. One may not add the
dark companions of two point baryonic masses, say; for, $\lambda$
is not proportional to M but rather to its square root. b) The
dark companion of a localized point mass is not itself localized.
Certainly, more accurate rotation speeds, specially in orbits
outside the plane of the galaxy will be helpful.

\end{document}